\documentclass[journal=jacsat,manuscript=article]{achemso}

\usepackage[version=3]{mhchem}

\usepackage{placeins}
\usepackage{booktabs}
\usepackage{multirow}
\usepackage{tabularx}
\usepackage{makecell}
\usepackage{rotating}
\usepackage{arydshln}
\usepackage{threeparttable}
\usepackage{hyperref}

\usepackage[T1]{fontenc}
\usepackage{amsmath}

\author{Jens Wagner}
\author{Thomas Specht}
\author{Hans Hasse}
\author{Fabian Jirasek}
\email{fabian.jirasek@rptu.de}
\phone{+49 (0)631 - 205 4685}
\affiliation[Unknown University]
{Laboratory of Engineering Thermodynamics (LTD), RPTU Kaiserslautern, Germany}

\title[An \textsf{achemso} demo]
  {Composition-Dependent Self-Diffusion Coefficients in Liquid Mixtures from Hybrid Machine Learning}

\begin{document}

\begin{abstract}

Self-diffusion coefficients are key descriptors of molecular mobility, yet experimental data remain scarce, highlighting the need for reliable prediction methods.
In previous work, we introduced the hybrid Enhanced Stokes--Einstein~(ESE) model, which advanced the state of the art in the physically consistent prediction of self-diffusion coefficients of solutes at infinite dilution in pure solvents by integrating the Stokes--Einstein equation with machine learning~(ML).
Here, we extend this approach to concentration-dependent self-diffusion coefficients and multicomponent solvents with HADES. This hybrid architecture leverages a deep-set neural network to connect pure-component and mixture prediction within a single framework.
HADES predicts self-diffusion coefficients in liquid mixtures with any number of components at any composition and temperature.
The only required inputs are SMILES-encoded molecular structures of the components and the pure-component viscosities, making the method broadly applicable.
Trained and evaluated on a comprehensive dataset of 2526 data points for 600 systems, HADES significantly outperforms benchmark prediction methods.
The trained model and its source code are fully disclosed, and the application is available via an interactive website \href{https://ml-prop.mv.rptu.de/}{\mbox{https://ml-prop.mv.rptu.de/}}.

\end{abstract}

\section{Introduction}

Self-diffusion coefficients in liquids quantify molecular mobility and are essential in many areas of science and engineering, including mixture characterization~\cite{Bellaire2020, Phuong2023}.
They also form the basis for modeling mutual diffusion, which is central to describing transport processes~\cite{Wesselingh2000, Bellaire2022b}.
However, experimental data on self-diffusion coefficients remain scarce~\cite{Gromann2022}.
Furthermore, they are often only available for solutes at infinite dilution in pure solvents, while concentration-dependent data and data for multicomponent mixtures are generally missing. 
Consequently, reliable prediction methods are needed to close these gaps. Existing methods~\cite{Taylor1993, Wilke1955, Reddy1967, Tyn1975, Evans2013, Evans2018, Eyring1941, Holmes1962, Tang1965, Leffler1970, Dullien1985, Carman1956, Li2001, Krishna2005, Liu2011, Khajeh2011, Abbasi2014, Mariani2020, Aniceto2021, Allers2022, Aniceto2024, Dias2024, Gromann2022, Romero2025, Romero2026} for predicting self-diffusion coefficients often only have a narrow scope and require input that may not be available.

To address these limitations, we developed HADES, a hybrid model for predicting self-diffusion coefficients in liquid mixtures as a function of temperature, covering mixtures with any number of components over the entire concentration range --- from infinite dilution to pure components.
HADES builds on previous work predicting self-diffusion coefficients at infinite dilution in pure solvents, the Enhanced Stokes--Einstein~(ESE) model~\cite{Wagner2026b}, which we augment and extend to concentration-dependent data and multicomponent mixtures.
HADES also lays the basis for a future extension to mutual diffusion.

\begin{figure}
\centering
    \includegraphics[width=14.25cm]{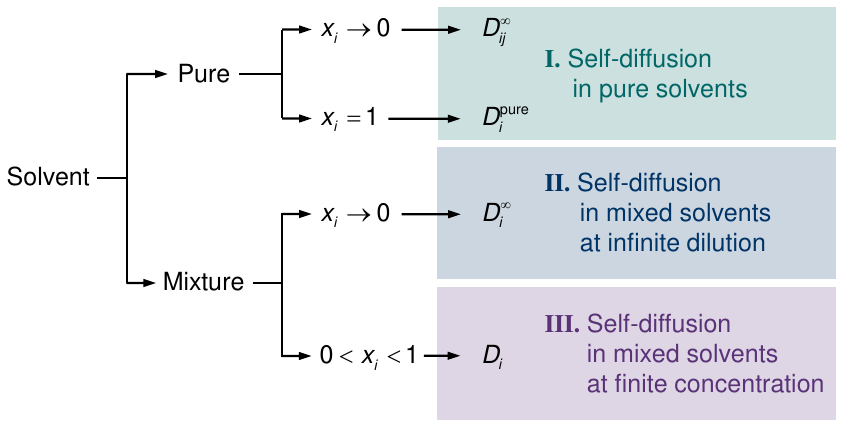}
    \caption{Classification of self-diffusion coefficients into three classes I - III used throughout this work, based on the solvent composition and the mole fraction~$x_i$ of the diffusing component~$i$.}
    \label{fig_Cla}
\end{figure}

The self-diffusion coefficient~$D_i$ describes the Brownian motion of a molecule of component~$i$ and is defined for pure components as well as for mixtures.
At infinite dilution in a binary mixture, the self-diffusion coefficient of a solute~$i$ in a pure solvent~$j$, $D^\infty_{ij}$, is identical to the corresponding Fickian and Maxwell--Stefan mutual diffusion coefficients and plays a central role in many semi-empirical approaches for predicting mutual diffusion in mixtures~\cite{Vignes1966, Kooijman1991, Darken1948, Krishna2005}.
At finite concentration, component~$i$ itself contributes to the mixed solvent in which its molecules diffuse.
In the opposite limit of a pure component, where $i$ acts as its own pure solvent, the self-diffusion coefficient reduces to the pure-component value~$D^\mathrm{pure}_i$.
In the present work, we classify self-diffusion and the corresponding self-diffusion coefficients into three classes according to the composition of the solvent and the concentration of the diffusing component~$i$, as shown in Figure~\ref{fig_Cla}, in accordance with the available experimental data and the scope of the corresponding benchmark models discussed below.

\FloatBarrier

While many methods for predicting self-diffusion coefficients in pure solvents exist~\cite{Taylor1993, Wilke1955, Reddy1967, Tyn1975, Evans2013, Evans2018, Khajeh2011, Abbasi2014, Mariani2020, Aniceto2021, Aniceto2024, Dias2024, Gromann2022, Romero2025, Romero2026, Wagner2026b}, there are fewer methods for predicting the concentration dependence of the self-diffusion coefficient in mixtures~\cite{Eyring1941, Holmes1962, Tang1965, Leffler1970, Dullien1985, Carman1956, Li2001, Krishna2005, Liu2011, Allers2022}.
These methods can be classified as follows: a)~Physical models and their empirical extensions~\cite{Taylor1993, Wilke1955, Reddy1967, Tyn1975, Evans2013, Evans2018, Eyring1941, Holmes1962, Tang1965, Leffler1970, Dullien1985, Carman1956, Li2001, Krishna2005, Liu2011}; b)~Data-driven models, based either on quantitative structure-property relationships~(QSPR) or machine learning~(ML)~\cite{Khajeh2011, Abbasi2014, Aniceto2021, Allers2022, Aniceto2024, Dias2024}; and c)~Hybrid models, combining ML with physically motivated models~\cite{Mariani2020, Gromann2022, Romero2025, Romero2026, Wagner2026b}.
Suitably developed hybrid models can combine the robustness of physical models with the flexibility of data-driven models~\cite{Jirasek2023b, Hasse2026}.
For a discussion of models for predicting self-diffusion coefficients in pure solvents~\cite{Taylor1993, Wilke1955, Reddy1967, Tyn1975, Evans2013, Evans2018, Khajeh2011, Abbasi2014, Mariani2020, Aniceto2021, Aniceto2024, Dias2024, Gromann2022, Romero2025, Romero2026}, we refer the reader to our recent work on the ESE model~\cite{Wagner2026b}.

For mixtures, Maxwell--Stefan~(MS) theory relates the self-diffusion coefficient of a solute~$i$ at infinite dilution in a mixed solvent, $D^\infty_i$, to the corresponding coefficients in the pure solvents, $D^\infty_{ik}$.
This relationship, given by Eq.~\eqref{eq_MS}, predicts the dependence of $D^\infty_i$ on the solvent composition:
\begin{equation}
D_i^\infty =
\frac{1}{\displaystyle\sum\limits_{\substack{k=1 \\ k \neq i}}^{N}
\frac{x_k}{D^\infty_{ik}}}.
\label{eq_MS}
\end{equation}
Here, $N$ is the total number of components in the mixture and $x_k$ is the mole fraction of solvent component~$k$.
Eq.~\eqref{eq_MS} follows directly from MS theory without assuming an ideal mixture. However, it does not account for the non-ideality of the solvent viscosity~\cite{Taylor1993}. In practice, good predictions for $D^\infty_i$ are generally obtained only for ideal or mildly non-ideal mixtures; for strongly non-ideal systems, predictions may deviate considerably from experimental data~\cite{Safi2006, Safi2010}.
Several empirical correlations have therefore been proposed~\cite{Eyring1941, Holmes1962, Tang1965, Leffler1970, Dullien1985} that introduce mixture viscosity as an additional composition-dependent input, improving predictions in some cases~\cite{Safi2006, Safi2010}.

Beyond the infinite-dilution limit, several physical and empirical models~\cite{Carman1956, Li2001, Krishna2005, Liu2011} have been proposed to predict the composition dependence of self-diffusion coefficients in mixtures at finite concentrations,~$D_i$.
In a recent comparison~\cite{GuevaraCarrion2016} with available experimental data for $D_i$, the model of Liu et al.~\cite{Liu2011}, which consistently extends the MS model (Eq.~\eqref{eq_MS}) to finite concentrations by accounting for the contribution of the diffusing component~$i$ to the solvent through the pure-component self-diffusion coefficient~$D_i^\mathrm{pure}$, was shown to give the best overall performance:
\begin{equation}
D_i =
\frac{1}{\displaystyle\sum\limits_{k=1}^{N}
\frac{x_k}{D^\infty_{ik}}}, \qquad \mathrm{with}\ D^{\infty}_{ii} = D^\mathrm{pure}_{i}.
\label{eq_Liu}
\end{equation}

Despite these efforts to model the composition dependence of self-diffusion coefficients in mixed solvents, the predictive accuracy of these physical and empirical models remains limited, with substantial deviations from experimental data reported even for weakly non-ideal mixtures~\cite{Safi2006, Safi2010, Bellaire2020}.

ML models provide an alternative to these approaches~\cite{Jirasek2021}.
However, only one ML model for predicting finite-concentration self-diffusion coefficients in binary mixtures has been proposed~\cite{Allers2022}.
It uses component- and mixture-specific descriptors, including critical-component data and association energies, as input.
As this model is fully data-driven, it lacks built-in physical constraints, such as monotonic increases in predicted self-diffusion coefficients with temperature, invariance to permutations of mixture components, and smooth variation with composition, which may produce unphysical results, particularly when applied outside the range of the experimental data used for training.

In addition to the limitations discussed above, the practical applicability of existing models is restricted by the availability of the required input data.
Physical and empirical approaches rely on self-diffusion coefficient data for $D^\infty_{ik}$ and $D_i^\mathrm{pure}$, which are available experimentally only for a small number of systems~\cite{Gromann2022}.
Extended and data-driven models require additional inputs, such as mixture viscosities and mixture-specific parameters, which further narrow their applicability.

Taken together, existing approaches for predicting self-diffusion coefficients in mixtures remain limited by prediction accuracy, input-data availability, restricted scope, or a lack of built-in physical consistency.

To address these limitations, we build on our hybrid ESE~\cite{Wagner2026b} framework to predict self-diffusion coefficients across the full composition range, from infinite dilution to pure components, introducing HADES --- the Hybrid Architecture for consistent Diffusion-coefficient Estimation across Solvents.
The resulting unified model predicts self-diffusion coefficients in liquid mixtures with arbitrary numbers of components at any composition.
Specifically, for class II, cf. Fig.~\ref{fig_Cla}, rather than enforcing Eq.~\eqref{eq_MS} as a hard constraint, HADES learns the composition dependence of $D^\infty_i$ directly from data: a deliberate modeling choice that sacrifices formal consistency with MS theory in favor of improved predictive accuracy, particularly for non-ideal mixtures.
HADES was trained jointly on self-diffusion coefficients in pure and mixed solvents to provide continuous predictions across composition, requiring only the molecular structures of the components (provided as SMILES) and the pure-component viscosities.

The remainder of this paper is structured as follows:
First, we briefly review the ESE model as the starting point for developing HADES.
Then, the HADES architecture, the experimental database, and the training and evaluation procedure are described.
Finally, HADES is systematically evaluated and benchmarked against the best available prediction models for the respective classes of self-diffusion coefficients (cf. Figure~\ref{fig_Cla}), namely the ESE~model~\cite{Wagner2026b} for class~I, the MS~model~(Eq.~\eqref{eq_MS}) for class~II, and the Liu model~\cite{Liu2011}~(Eq.~\eqref{eq_Liu}) for class~III. 

\section{Starting-Point: The ESE Model}

The Stokes--Einstein~(SE) equation~\cite{Einstein1905} provides the physical foundation for predicting self-diffusion coefficients~$D_i$ in a liquid solvent.
It was derived considering the motion of a hard sphere (diffusing molecule~$i$) in a continuous fluid (solvent) yielding
\begin{equation}
D_{i}^\mathrm{SE} = \frac{k_\mathrm{B} T}{6 \pi \eta r_i},
\label{eq_SE}
\end{equation}
where $k_\mathrm{B}$ is the Boltzmann constant, $T$ is the temperature, $\eta$ is the dynamic viscosity of the solvent at the temperature $T$, and $r_i$ is the effective radius of the solute~$i$, which can be estimated from the molar mass of the solute $M_i$, its specific density $\rho_i$, and the Avogadro constant $N_\mathrm{A}$ using:
\begin{equation}
r_i = \sqrt[3]{\frac{3 M_i}{4 \pi \rho_i N_\mathrm{A}}f},
\label{eq_ri}
\end{equation}
where $f$ is an empirical correction factor that can be related to the packing fraction and is often reported to be in the range of $f = 0.6\,\,\ldots\,\,0.8$.
For predictive applications in liquids, $f = 0.64$ is frequently used as a default~\cite{Evans2013}.

In previous work~\cite{Wagner2026b}, we introduced the hybrid Enhanced Stokes--Einstein~(ESE) model for the physically consistent prediction of self-diffusion coefficients of a solute~$i$ at infinite dilution in pure solvents~$j$, $D^{\infty, \mathrm{ESE}}_{ij}$, by integrating the Stokes--Einstein equation with a neural network (NN).
In the ESE model, the prediction of the SE model, $D^{\infty, \mathrm{SE}}_{ij}$, is refined using a system-specific scaling factor, $b_{ij}$, to enhance its predictive capacity.
This scaling factor is generated by the NN leveraging simple molecular descriptors of both the solute and solvent, $\mathbf{X}_i$ and $\mathbf{X}_j$, respectively, which are automatically derived from the SMILES~\cite{Weininger1988} strings of the components using the open-source tool RDKit~\cite{RDKit2024}.
The architecture of the ESE model naturally accommodates the prediction of self-diffusion coefficients in pure components by choosing $j = i$, such that $D^\mathrm{pure,ESE}_{i} = D^{\infty, \mathrm{ESE}}_{ii}$.

Importantly, the NN architecture is explicitly constrained to preserve the physical principles encoded in the SE model, ensuring that the ESE model delivers consistent predictions across temperatures while benefiting from the flexibility of the ML algorithm.
Beyond the SMILES strings, the only required physico-chemical input is the solvent viscosity, $\eta_j$, at the temperature of interest $T$, which is also needed for applying the SE model itself, making the ESE model broadly applicable.
We benchmarked the ESE model against several high-performing previously published models~\cite{Evans2018, Romero2025, Romero2026}, demonstrating its broader applicability and superior predictive performance across a wide range of systems and temperatures.
For further details on the ESE model, see the following sections and the original work~\cite{Wagner2026b}.

In this work, we extend the ESE model from self-diffusion coefficients in pure solvents to predicting concentration-dependent self-diffusion coefficients in binary, ternary, and multicomponent systems.
Thereby, the initial ESE model was also updated, so that we recommend using HADES instead of ESE in future applications.

We distinguish here between the diffusing component~$i$ and its solvent.
In ESE, this solvent is always pure, whereas it can be a mixture in HADES.
Note that this mixed solvent may also contain the component~$i$ for which the self-diffusion is measured, namely, if $i$ is not considered at infinite dilution.
To account for multiple components~$k$ in the solvent, HADES incorporates an equivariant deep-set model~(DSM)~\cite{Zaheer2017}, which aggregates their individual contributions~$\mathbf{X}_k$ into $\mathbf{X}_\mathrm{solv}$, allowing predictions for solutes at any concentration and systems with any number of components, while guaranteeing results invariant to their permutation.
As HADES at its core relies on the SE model, also the viscosity of the mixed solvent is needed, for which logarithmic mixing of the component viscosities is assumed throughout this work.
Hence, the only inputs required for HADES are the SMILES strings of all components as well as the pure-component viscosities of the solvent components~$k$ at the temperature~$T$ of interest. 

\section{Methods}

\subsection{Model Architecture}
\label{sec_ma}

\begin{figure}
\centering
    \includegraphics[width=15cm]{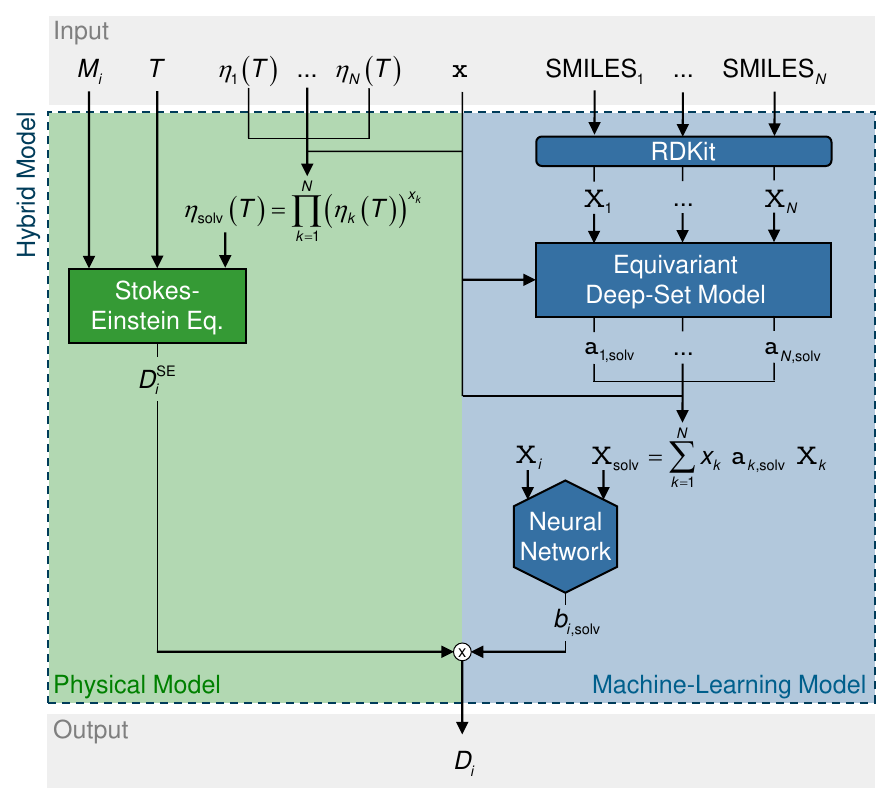}
    \caption{Schematic overview of the HADES architecture for predicting self-diffusion coefficients~$D_{i}$ at any concentration in pure and mixed solvents. Details on the model architecture are given in Section~\nameref{sec_ma}.}
    \label{fig_Met}
\end{figure}

Figure~\ref{fig_Met} provides a schematic overview of the architecture of HADES.
The gray box shows the input.
In the scheme, the index~$i$ refers to the component for which the self-diffusion coefficient is calculated.
The green box shows the steps for applying the SE model, i.e., the physical part of the model.
To apply the SE equation, first the viscosity of the mixture (the solvent) $\eta_\mathrm{solv}$ is calculated from the pure-component viscosities~$\eta_k$ and the mole fractions of the components~$x_k$ using:
\begin{equation}
\eta_\mathrm{solv}(T) = \prod^{N}_{k=1} \left( \eta_k(T) \right) ^{x_k}.
\label{eq_IV}
\end{equation}

The self-diffusion coefficient $D^\mathrm{SE}_i$ is then calculated from Eqs.~\eqref{eq_SE} and~\eqref{eq_ri} using the molar mass $M_i$ and the temperature $T$.
To avoid additional input, we used the default value $f = 0.64$ throughout and set the specific density of component $i$ to $\rho_i = 1050$~kg~m$^{-3}$ for all components.
These choices (as well as the choice of Eq.~\eqref{eq_IV} for calculating the solvent viscosity) were made based on preliminary tests, which showed that their influence is small~\cite{Wagner2026b}, as the result of the physical model, $D^\mathrm{SE}_i$, is corrected by a factor $b_{i, \mathrm{solv}}$ to obtain the final results of the self-diffusion coefficient $D_i$:
\begin{equation}
D_i = D_i^{\mathrm{SE}} \cdot b_{i, \mathrm{solv}}.
\label{eq_HADES}
\end{equation}
The factor $b_{i, \mathrm{solv}}$ is calculated in the ML part of the model (blue box in Figure~\ref{fig_Met}), which is trained to compensate for errors in the physical part and explains that the strongly simplifying assumptions described above have no negative consequences.

The ML part of HADES takes as input the SMILES strings and the mole fractions~$x_k$ of all $N$ mixture components. The SMILES are first processed in the open-source toolkit RDKit~\cite{RDKit2024} to automatically generate the molecular descriptor vectors~$\mathbf{X}_k$.
The molecular descriptor set used to represent the mixture components in this work, which is listed in Table~\ref{tab_x}, is the same as in the ESE model~\cite{Wagner2026b} and was not varied or optimized during HADES development. The molecular descriptor vectors~$\mathbf{X}_k$ are used together with the mole fractions~$x_k$ of the respective components as inputs to an equivariant DSM~\cite{Zaheer2017}.
Unlike classical neural networks, this architecture processes inputs in parallel, enabling the model to capture interactions between them (here, between the solvent components) while ensuring that the outputs are independent of input order and can still be assigned to the corresponding inputs~\cite{Soelch2019, Wagstaff2022}.
For every component, this architecture produces a solvent-specific vector $\mathbf{a}_{k, \mathrm{solv}}$ of the same dimension as $\mathbf{X}_k$, which weights the component’s molecular descriptors according to their influence on the self-diffusion in the considered solvent.
The component-specific descriptors are then weighted by the respective mole fractions $x_k$ and combined to obtain the solvent’s molecular descriptor $\mathbf{X}_\mathrm{solv}$:
\begin{equation}
\mathbf{X}_\mathrm{solv} = \sum_{k=1}^{N} x_k \ \mathbf{a}_{k, \mathrm{solv}} \ \mathbf{X}_k.
\label{eq_sol}
\end{equation}
The sum aggregation in Eq.~\eqref{eq_sol} guarantees that $\mathbf{X}_\mathrm{solv}$ is invariant to any permutation of the solvent components and that HADES produces predictions that vary smoothly with composition for solvents with any number of components.

\begin{table}
\centering
\caption{Molecular descriptors included in the vector~$\mathbf{X}$ describing the solute and solvent components as used in the HADES model.}
\label{tab_x}
    \begin{tabular}{ll}
        \toprule
        Label & Molecular descriptor \\
        \midrule
        $M$ & Molar mass in kg~mol$^{-1}$\\
        $R$ & Boolean variable indicating presence of molecular ring structures\\
        $r_\mathrm{Het}$ & Ratio of number of heteroatoms to non-hydrogen atoms\\
        $r_\mathrm{Hal}$ & Ratio of number of halogen atoms to non-hydrogen atoms\\
        $r_\mathrm{Acc}$ & Ratio of number of hydrogen-bond acceptors to non-hydrogen atoms\\
        $r_\mathrm{Don}$ & Ratio of number of hydrogen-bond donors to non-hydrogen atoms\\
        \bottomrule
    \end{tabular}
\end{table}
\FloatBarrier

Analogously to the ESE model, the solvent descriptor, $\mathbf{X}_\mathrm{solv}$, and the descriptor of the component for which the self-diffusion is to be predicted, $\mathbf{X}_i$, are fed to a NN that yields the system-specific scaling factor $b_{i, \mathrm{solv}}$ for the correction of the SE prediction, cf. Eq.~\eqref{eq_HADES}.
The NN is restricted to produce strictly positive outputs for $b_{i, \mathrm{solv}}$.
Furthermore, because all descriptors used in the ML model are structural and temperature-independent, the temperature dependence of the SE model is preserved.

The DSM architecture consists of an input layer, a deep-set layer with sum aggregation~\cite{Zaheer2017, Soelch2019}, and an output layer, each with 32 nodes.
The Sigmoid Linear Unit (SiLU) activation function was used.
Details on the hyperparameter optimization are given below.
The NN architecture was adopted from our previous work~\cite{Wagner2026b}, with the only modification being that, as in the DSM, the inter-layer activation function was SiLU.

\subsection{Experimental Database}

This work is focused on organic molecular substances with molar masses below 1000~g~mol$^{-1}$ that contain no heavier atoms than chlorine.
Additionally, we also consider water.
The resulting scope is consistent with the scope of the ESE model~\cite{Wagner2026b}.
The database used in this work contains $N = 2526$ experimental data points of self-diffusion coefficients comprising 215~components for which the self-diffusion is reported and 48 components present in the corresponding solvents.
Table~\ref{tab_db} gives an overview.
A complete list of all diffusing components and solvent components included in the database is provided in the Supporting Information.
The dynamic viscosities of all solvent components at the respective temperatures were obtained from the Dortmund Data Bank~(DDB)~\cite{DDB2024}.

\begin{sidewaystable}
\centering
\caption{Overview of the experimental database and corresponding benchmark models for the different classes of self-diffusion coefficients (cf. Figure~\ref{fig_Cla}).}
\label{tab_db}
\begin{tabular}{lllccccccc}
    \toprule
    Class
    & \makecell{Self-diffusion\\coefficient}
    & System
    & \multicolumn{6}{c}{Database}
    & Benchmark \\
    \cmidrule(lr){4-9}
    &
    &
    & Systems
    & \makecell{Diffusing\\components}
    & Solvents
    & \makecell{Temperature\\range}
    & \makecell{Data\\points}
    & Sources
    & \\
    \midrule
    \multirow{2}{*}{I}
    & $D_{ij}^\infty$ & Binary & 474 & 209 & 42 & 273~K -- 363~K & 1011 & \citenum{Wagner2026b} & \multirow{2}{*}{ESE model~\cite{Wagner2026b}} \\
    & $D_i^\mathrm{pure}$ & Pure & 19 & 19 & 19 & 273~K -- 363~K & 56 & \citenum{DDB2024} & \\

    \multirow{2}{*}{II}
    & \multirow{2}{*}{$D_i^\infty$}
    & Ternary & 95 & 11 & 23 & 298~K & 713 & \multirow{2}{*}{\citenum{Safi2006, Safi2007, Safi2008, Safi2010}} & \multirow{2}{*}{\makecell[c]{MS model\\(Eq.~\eqref{eq_MS})}} \\
    & & Quaternary & 4 & 1 & 4 & 298~K & 37 &  & \\

    \multirow{2}{*}{III}
    & \multirow{2}{*}{$D_i$}
    & Binary & 18 & 17 & 18 & 293~K -- 358~K & 504 & \multirow{2}{*}{\citenum{DDB2024}} & \multirow{2}{*}{\makecell[c]{Liu model~\cite{Liu2011}\\(Eq.~\eqref{eq_Liu})}} \\
    & & Ternary & 3 & 5 & 3 & 298~K & 205 &  & \\
    \bottomrule
\end{tabular}
\end{sidewaystable}
\FloatBarrier

The experimental data for $D^\infty_{ij}$ are the same as those used in our previous work on the ESE model~\cite{Wagner2026b}, mainly comprising data collected from the DDB in previous works~\cite{Gromann2022, Romero2025} and measurements performed in our own laboratories~\cite{Romero2025, Romero2026, Wagner2025, Bellaire2022, Mross2024}.
Data for $D_i^\mathrm{pure}$ and $D_i$ were analogously obtained from the DDB~\cite{DDB2024}.
For $D_i^\infty$, available experimental data in the literature are limited to systems with aromatic diffusing components and are reported in four sources~\cite{Safi2006,Safi2007,Safi2008,Safi2010}, all of which were considered here.

\subsection{Training and Evaluation}
\label{sec_tae}

HADES was trained jointly on data for $D_{ij}^\infty$, $D_i^\infty$, $D_i^\mathrm{pure}$, and $D_i$ (cf. Figure~\ref{fig_Cla}).
This enables the model to learn consistent trends across different classes.
We trained the model using $K$-fold cross-validation~(CV)~\cite{Stone1974} with solvent-wise data splits, where $K = 92$ equals the number of distinct solvents in our database.
In each fold, all data associated with one solvent (independent of solvent component concentrations and temperature) were held back as test data.
The remaining data were randomly split point-wise into 80\% training data and 20\% validation data.
This procedure was repeated until predictions for all data points in the database were obtained.

During training, the weights of the DSM and the NN were optimized simultaneously to predict $b_{i, \mathrm{solv}}$ (cf. Eq.~\eqref{eq_HADES}) by minimizing the mean squared relative error~(MSRE), which is defined as the average, over all considered data points, of the squared relative error~(SRE) between the predicted and experimental values:
\begin{equation} \mathrm{SRE} = \left( \frac{D^{\mathrm{pred}} - D^{\mathrm{exp}}}{D^{\mathrm{exp}}} \right)^2 .
\label{eq_SRE}
\end{equation}

Weight updates were performed using the AdamW~\cite{Loshchilov2017} optimizer.
The learning rate was reduced by a factor of 0.1 after 25 epochs without improvement in the validation loss. Early stopping was applied when no improvement was observed for 50 consecutive epochs, and the model with the lowest validation loss was used for evaluation on the corresponding test set.

Hyperparameters were optimized using a grid search approach based on the validation loss.
The final hyperparameters, together with a sensitivity analysis on their influence on the validation loss, are reported in the Supporting Information.

In addition to SRE, the predictive performance was also assessed using the absolute relative error~(ARE):
\begin{equation} \mathrm{ARE} = \left| \frac{D^{\mathrm{pred}} - D^{\mathrm{exp}}}{D^{\mathrm{exp}}} \right|,
\label{eq_ARE}
\end{equation}
and its average over all considered data points, the mean absolute relative error~(MARE).

While the solvent-wise CV was used to assess predictive capability, we also provide a ’final version’ of HADES.
We obtained this final version using an ensemble of ten models~\cite{Dietterich2000}, trained by generating ten independent random point-wise splits of the full dataset (95\% training, 5\% validation).
For each split, an individual model was initialized with a different random seed and trained separately.
The ensemble prediction is obtained as the average of the individual models' results.
The final HADES ensemble is available on Zenodo (\href{https://doi.org/10.5281/zenodo.21979762}{\mbox{https://doi.org/10.5281/zenodo.21979762}}) and through the interactive MLPROP~\cite{Hoffmann2025} (\href{https://ml-prop.mv.rptu.de/}{\mbox{https://ml-prop.mv.rptu.de/}}). 

Training was executed on a CPU node of the HPC Elwetrisch (dual Intel Xeon Silver 4410T, 64 GB RAM), using a single CPU core.
Models, training, and evaluation routines were implemented in Python 3.12.7 using PyTorch 2.2.1~\cite{Paszke2019}.
Typical training times per fold ranged from 8 to 10 min.

To enable direct comparison with HADES in predicting self-diffusion coefficients in pure solvents, we retrained and reevaluated the ESE model using the same procedure, architecture, and hyperparameter settings as described in the original work~\cite{Wagner2026b}, but adopting the solvent-wise CV and database for $D_{ij}^\infty$ and $D_i^\mathrm{pure}$ from this work, thereby aligning the ESE training and evaluation with that of HADES.
All ESE results reported here were obtained with this retrained version. 

\section{Results and Discussion}

In the following, we evaluate the performance of HADES for predicting self-diffusion coefficients in pure components and mixtures using test data for unseen solvents (see Section~\nameref{sec_tae}).
We also compare HADES results to benchmark models.
There are two classes of benchmarks: a) physics-based models that were directly taken from the literature, and b) the hybrid ESE~model, which was the only data-driven or ML-based model available in a form that enabled direct benchmarking and was re-parameterized in the present work to enable a fair and unbiased comparison.

A direct comparison, as carried out between HADES and ESE, is unfortunately not possible for the physics-based models, since these models require input information that differs from that required by HADES.
In particular, they generally require experimental data for self-diffusion coefficients at infinite dilution and/or in pure components, but these are often unavailable. In these cases, ESE predictions were used to replace unavailable experimental input data.

\subsection{Self-Diffusion in Pure Solvents}
\label{sec_ESE}

\begin{figure}
\centering
    \includegraphics[width=16cm]{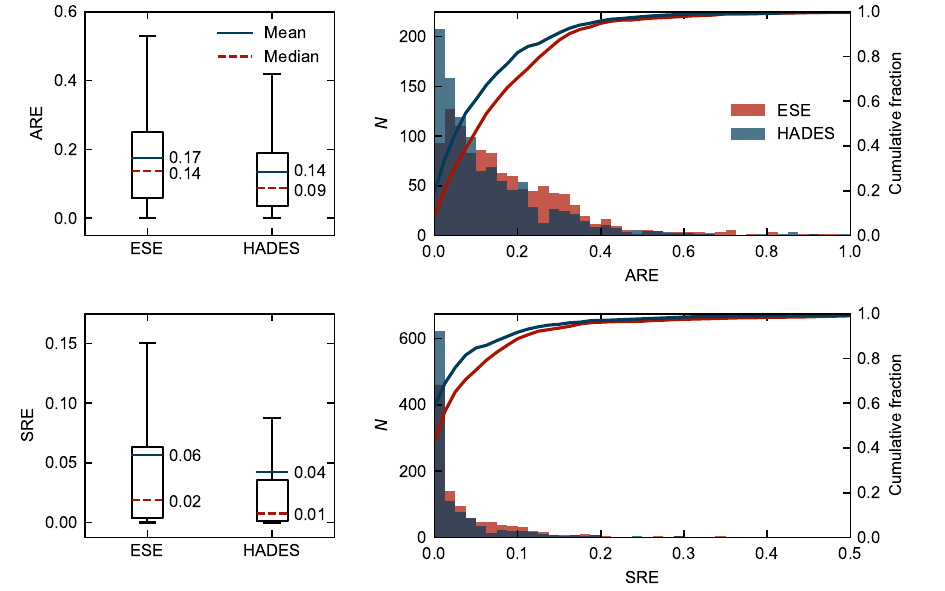}
    \caption{Absolute relative error~(ARE, top) and squared relative error~(SRE, bottom) of the predicted self-diffusion coefficients in pure solvents, $D^\infty_{ij}$ and $D^\mathrm{pure}_{i}$, using ESE and HADES.
    Left: Boxplots, with box height indicating the interquartile range (IQR) and whiskers 1.5 times the IQR.
    Right: Histograms (bars) with cumulative fractions (lines) showing the number of data points~$N$ predicted with a certain error.}
    \label{fig_ESE}
\end{figure}

Figure~\ref{fig_ESE} compares the self-diffusion coefficients in pure solvents, $D^\infty_{ij}$ and $D^\mathrm{pure}_{i}$, predicted by ESE and HADES with the experimental test data, in terms of the ARE (top) and SRE (bottom).
The results are presented as boxplots in the left panels and as histograms with the corresponding cumulative fractions in the right panels.
The results show that HADES consistently outperforms ESE in both the ARE and SRE.
As an example: while only approximately 21\% of the data points can be predicted with an $\mathrm{ARE}< 0.05$ with ESE, which is in the range of typical experimental uncertainties reported for $D^{\infty}_{ij}$~\cite{Gromann2022}, HADES achieves this level of accuracy for approximately 34\% of the data points.
\FloatBarrier

The improved predictive performance of HADES for self-diffusion coefficients in pure solvents is particularly noteworthy because the ESE model was specifically designed for this prediction task.
This finding indicates that the deep-set architecture chosen for HADES enables the model, unlike ESE, to exploit cross-correlations learned from self-diffusion data in mixed solvents, improving performance on the pure-solvent prediction task.

\subsection{Self-Diffusion in Mixed Solvents at Infinite Dilution}
\label{sec_MS}

\begin{figure}
\centering
    \includegraphics[width=16cm]{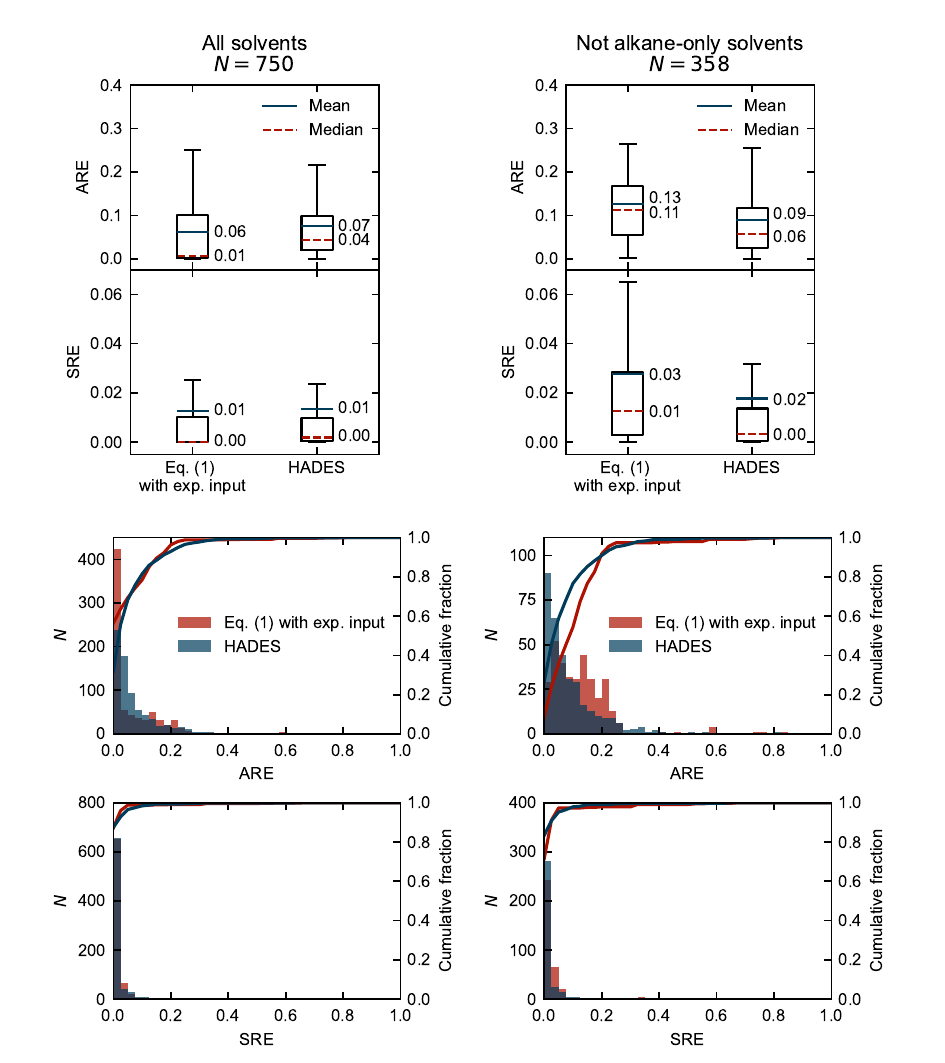}
    \caption{Absolute relative error~(ARE) and squared relative error~(SRE) of the predicted self-diffusion coefficients in mixed solvents at infinite dilution, $D^\infty_{i}$, using the MS model~(Eq.~\eqref{eq_MS}) and HADES for all systems~(left column) and for the subset of systems containing at least one non-alkane solvent component~(right column), with the respective number of experimental test data points~$N$.
    Top: Boxplots, with box height indicating the interquartile range (IQR) and whiskers 1.5 times the IQR.
    Bottom: Histograms (bars) with cumulative fractions (lines) showing the number of data points~$N$ predicted with a certain error.}
    \label{fig_MS}
\end{figure}

Figure~\ref{fig_MS} compares the self-diffusion coefficients in mixed solvents at infinite dilution, $D^\infty_{i}$, predicted by the MS model~(Eq.~\eqref{eq_MS}) and HADES with the experimental test data, in terms of the ARE and SRE.
The left column of Figure~\ref{fig_MS} shows the results for all data points.
Both the MS model and HADES yield generally low prediction errors and show comparable performance in predicting~$D^\infty_{i}$.
However, the MS model requires experimental self-diffusion coefficients in the pure solvent components, $D^\infty_{ij}$, as input, whereas HADES predicts $D^\infty_{i}$ without any experimental diffusion data.
The most notable difference is the lower median ARE of the MS model, which can be attributed to the larger fraction of data points predicted with very low ARE values, as reflected in the corresponding histogram.
This is reasonable, as most available $D_i^\infty$ data are for alkane solvent mixtures~\cite{Safi2006, Safi2007, Safi2008, Safi2010}, i.e., near-ideal systems, where Eq.~\eqref{eq_MS} is known to perform well in practice.

The right column of Figure~\ref{fig_MS} shows the corresponding results for the subset of systems containing at least one non-alkane solvent component.
For these systems, HADES shows clearly improved predictive performance over the MS model, indicating it is better suited to describe the composition dependence of self-diffusion coefficients in mixed solvents exhibiting non-ideal behavior.
\FloatBarrier

\begin{figure}
\centering
    \includegraphics[width=15cm]{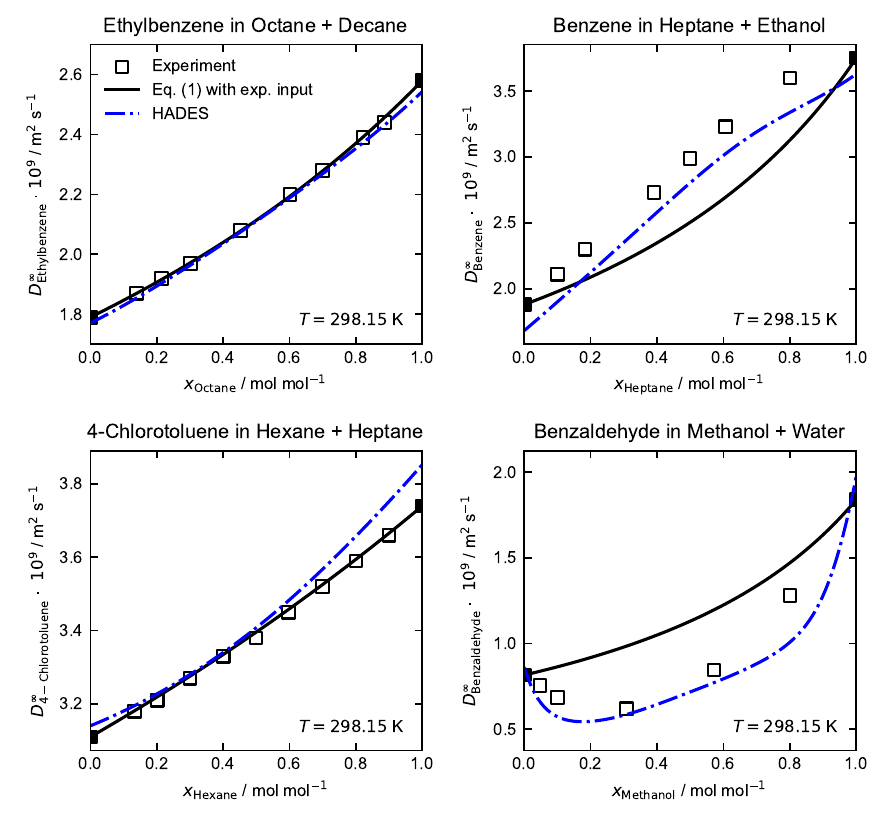}
    \caption{Self-diffusion coefficients in mixed solvents at infinite dilution, $D_i^\infty$, as a function of solvent composition for four exemplary test systems.
    Left column: systems with alkane mixtures as solvents.
    Right column: systems with at least one non-alkane solvent component.
    Unfilled symbols represent experimental test data, while lines denote predictions from the MS~(Eq.~\eqref{eq_MS}) and HADES models.
    Filled symbols indicate the experimental self-diffusion coefficients at infinite dilution in the pure solvent components used for application of Eq.~\eqref{eq_MS}.}
    \label{fig_MSII}
\end{figure}

Figure~\ref{fig_MSII} compares the self-diffusion coefficients in mixed solvents at infinite dilution, $D_i^\infty$, predicted by the MS model and HADES with the experimental test data as a function of solvent composition for four exemplary systems.

The left column shows systems with alkane-only solvent mixtures, where near-ideal behavior is expected.
For these systems, the MS model accurately predicts the experimental data across the full composition range, consistent with the theoretical basis of Eq.~\eqref{eq_MS}, which, while exact within MS theory, is known to perform well in practice precisely for near-ideal systems.
HADES performs equally well for ethylbenzene in (octane + decane), while it shows slight deviations for 4-chlorotoluene in (hexane + heptane).

The right column shows results for two non-ideal systems.
The MS model fails to describe the composition dependence of $D_i^\infty$ for both systems.
HADES, by contrast, accurately predicts the experimental data across the full composition range, capturing both the positive deviation from ideal behavior observed for benzene in (heptane + ethanol), and the pronounced negative deviation for benzaldehyde in (methanol + water).
These results directly illustrate the benefit of the modeling choice discussed above: by learning $D_i^\infty$ directly from data rather than enforcing Eq.~\eqref{eq_MS}, HADES gains the flexibility to describe non-ideal composition dependence that Eq.~\eqref{eq_MS} cannot capture.

In the examples shown in Figure~\ref{fig_MSII}, experimental values for pure-solvent coefficients $D_{ik}^\infty$ were used throughout for the predictions based on Eq.~\eqref{eq_MS}.
Such experimental data are often unavailable.
In principle, they could be replaced by model predictions, e.g., from HADES or ESE.
For ethylbenzene in (octane + decane) and benzaldehyde in (methanol + water), where HADES accurately predicts the pure-solvent coefficients, this substitution would have little effect.
For the other two systems, 4-chlorotoluene in (hexane + heptane) and benzene in (heptane + ethanol), predictions based on Eq.~\eqref{eq_MS} would deteriorate because deviations in the pure-solvent coefficients would carry over to the mixed-solvent results. 
\FloatBarrier

\begin{figure}
\centering
    \includegraphics[width=15cm]{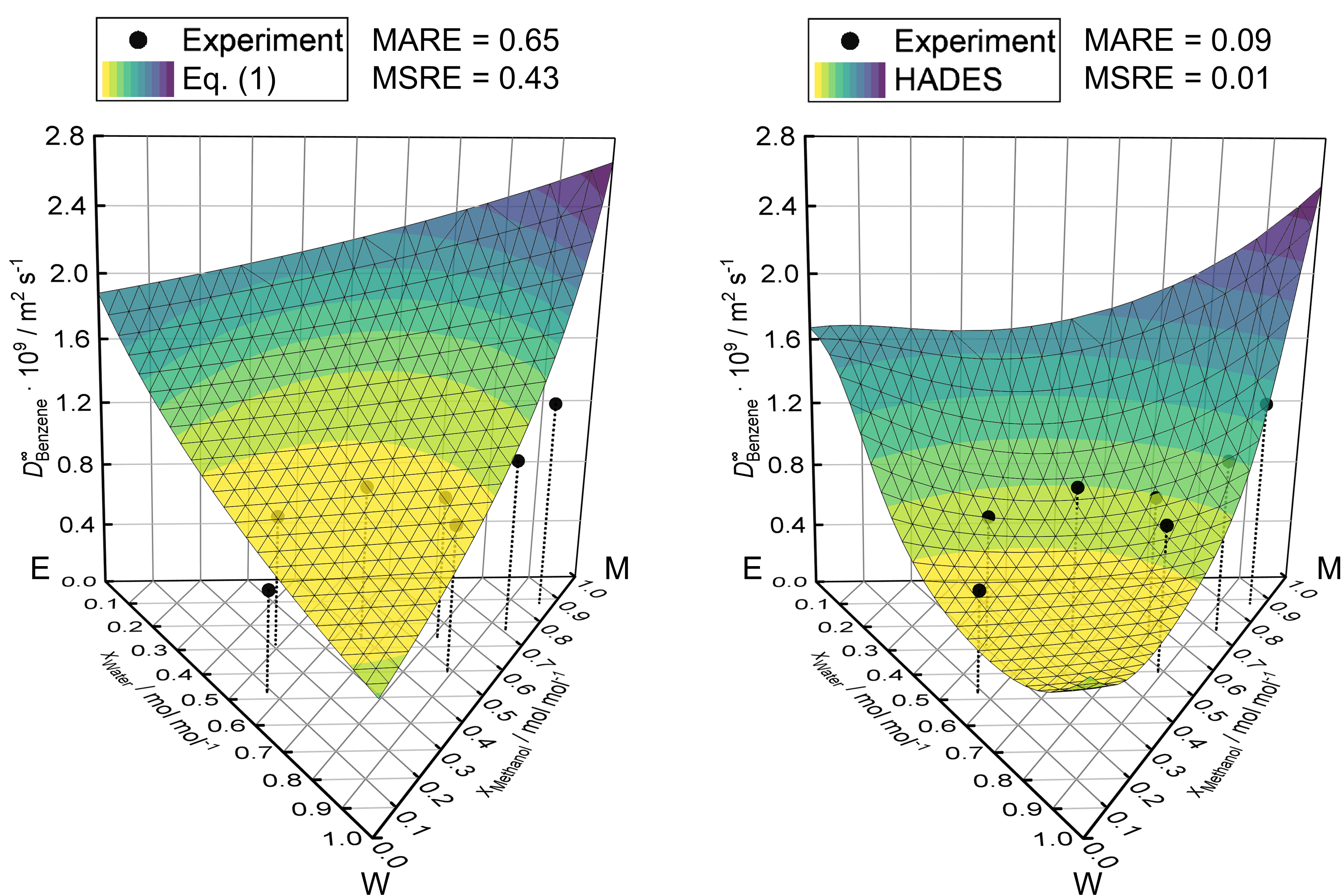}
    \caption{Self-diffusion coefficient of benzene at infinite dilution, $D_\mathrm{Benzene}^\infty$, as a function of solvent composition in a ternary mixture of methanol~(M), ethanol~(E), and water~(W) at $T=298.15$~K.
    Left: MS model~(Eq.~\eqref{eq_MS}) predictions compared with experimental test data; Eq.~\eqref{eq_MS} was applied using experimental self-diffusion coefficients at infinite dilution in the pure solvent components.
    Right: HADES predictions compared with experimental test data.
    Symbols represent experimental test data, while surfaces denote model predictions.}
    \label{fig_MSIII}
\end{figure}

Figure~\ref{fig_MSIII} compares the experimental test data for the self-diffusion coefficient of benzene at infinite dilution in a ternary solvent mixture of (methanol + ethanol + water) with predictions from Eq.~\eqref{eq_MS} and HADES.
While Eq.~\eqref{eq_MS} predicts an almost linear interpolation between the pure-solvent coefficients, HADES predicts strongly non-linear behavior. 
The MARE and MSRE values show that HADES predictions match the experimental test data for this non-ideal system much better than predictions from Eq.~\eqref{eq_MS}.
\FloatBarrier

Together, the results presented in Figures~\ref{fig_MS}~--~\ref{fig_MSIII} demonstrate that HADES predicts the composition dependence of self-diffusion coefficients at infinite dilution in mixed solvents exhibiting non-ideal behavior clearly better than Eq.~\eqref{eq_MS}, even in the rare case where experimental pure-solvent data are available for the application of Eq.~\eqref{eq_MS}.
Furthermore, the results shown in Figures~\ref{fig_MSII} and~\ref{fig_MSIII} demonstrate that, owing to its architecture, HADES provides consistent predictions over the full composition space for mixtures with any number of solvent components.

\subsection{Self-Diffusion in Mixed Solvents at Finite Concentration}
\label{sec_Liu}

\begin{figure}
\centering
    \includegraphics[width=16cm]{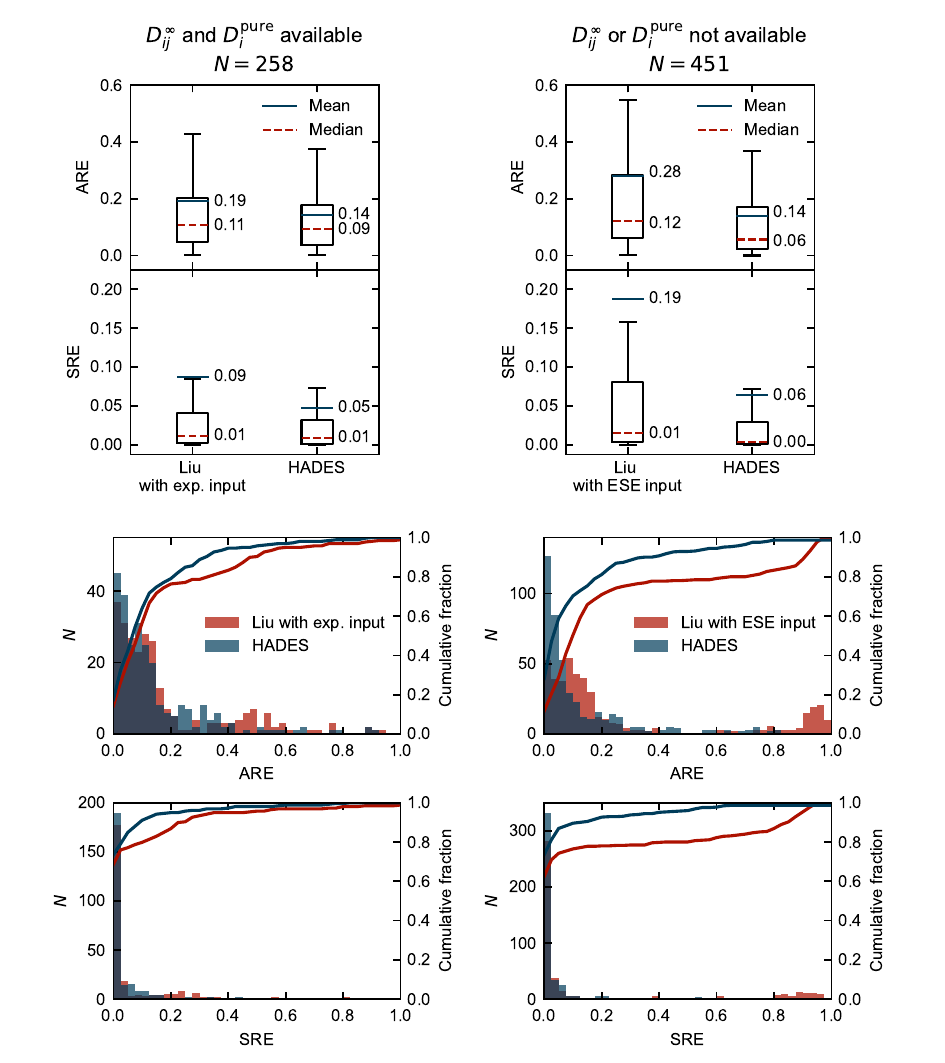}
    \caption{Absolute relative error~(ARE) and squared relative error~(SRE) of the predicted self-diffusion coefficients in mixed solvents at finite concentrations, $D_i$, using the Liu model~\cite{Liu2011}~(Eq.~\eqref{eq_Liu}) and HADES for mixtures for which experimental self-diffusion coefficients in the pure solvent components, $D_{ij}^\infty$ and $D_i^\mathrm{pure}$, are available for application of the Liu model~\cite{Liu2011} (left column) and mixtures for which such experimental data are unavailable and ESE predictions are used instead (right column), with the respective number of experimental test data points~$N$.
    Top: Boxplots, with box height indicating the interquartile range (IQR) and whiskers 1.5 times the IQR.
    Bottom: Histograms (bars) with cumulative fractions (lines) showing the number of data points~$N$ predicted with a certain error.}
    \label{fig_Liu}
\end{figure}

Figure~\ref{fig_Liu} compares the self-diffusion coefficients in mixed solvents at finite concentrations, $D_{i}$, predicted by the Liu model~\cite{Liu2011}~(Eq.~\eqref{eq_Liu}) and HADES with the experimental test data, in terms of the ARE and SRE.
The left column of Figure~\ref{fig_Liu} shows the results for mixtures for which the experimental self-diffusion coefficients in the respective pure solvents, $D^{\infty}_{ij}$ and $D^{\mathrm{pure}}_{i}$, are available for application of the Liu model~\cite{Liu2011}.
HADES performs markedly better than the benchmark model in both ARE and SRE, with the substantially lower MSRE (0.05 vs. 0.09) indicating, in particular, fewer strongly mispredicted data points.
As shown by the histogram, both models predict a comparable number of data points with small $\mathrm{ARE} < 0.15$; however, HADES is more robust, as reflected in the faster increase in the cumulative fraction at higher $\mathrm{ARE} > 0.2$.

The right column shows results for mixtures lacking experimental data for the pure solvent components, for which we used the re-parameterized version of ESE in applying the Liu model~\cite{Liu2011}.
For these mixtures, the Liu model~\cite{Liu2011} performs markedly worse. In contrast, HADES significantly outperforms the reference model and achieves an accuracy comparable to that obtained for mixtures with available experimental data in the pure solvent components.
\FloatBarrier

\begin{figure}
\centering
    \includegraphics[width=15cm]{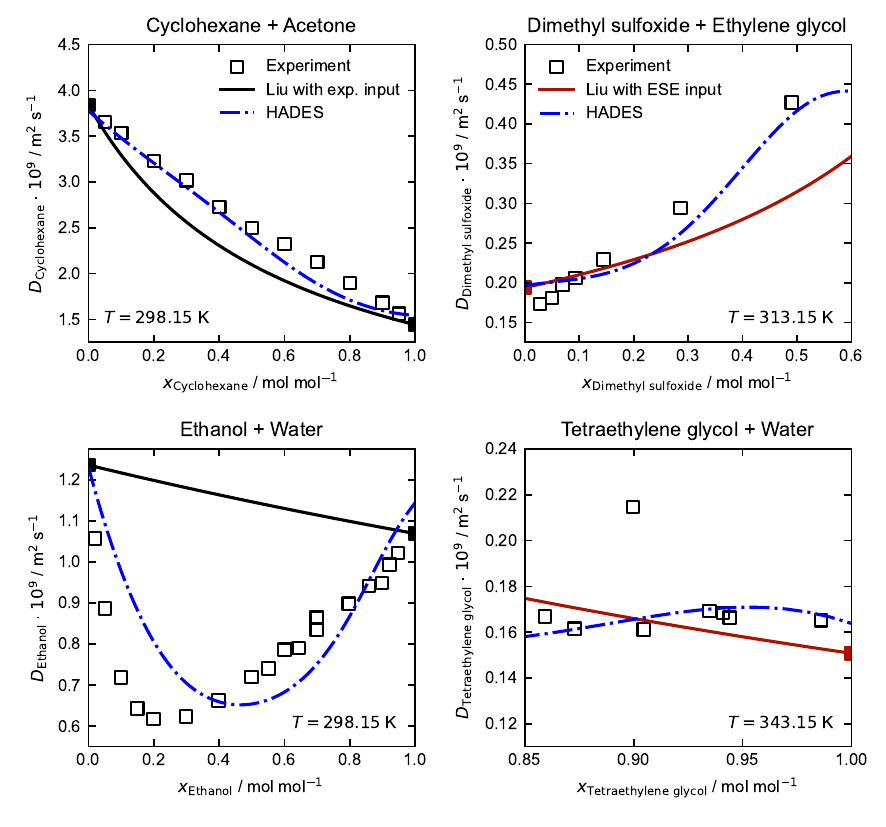}
    \caption{Self-diffusion coefficients in mixed solvents at finite concentrations, $D_i$, as a function of the mole fraction, $x_i$, for four exemplary binary mixtures.
    Left column: mixtures for which experimental self-diffusion coefficients in the pure solvent components, $D_{ij}^\infty$ and $D_i^\mathrm{pure}$, are available for application of the Liu model~\cite{Liu2011}.
    Right column: mixtures for which such experimental data are unavailable and ESE predictions are used instead.
    Unfilled symbols represent experimental test data, while lines denote predictions from the Liu~\cite{Liu2011} and HADES models.
    Filled symbols indicate the self-diffusion coefficients in the pure solvent components used for application of the Liu model~\cite{Liu2011}.}
    \label{fig_LiuII}
\end{figure}

Figure~\ref{fig_LiuII} compares the self-diffusion coefficients at finite concentrations, $D_i$, predicted by the Liu model~\cite{Liu2011} and HADES with experimental test data, as a function of the mole fraction~$x_i$, for four exemplary binary mixtures.
The left column of Figure~\ref{fig_LiuII} shows two mixtures for which experimental values of the self-diffusion coefficients in the pure solvent components are available.
For both mixtures, the Liu model~\cite{Liu2011} fails to capture the concentration-dependent behavior of the self-diffusion coefficient in the mixture, whereas HADES predicts it remarkably well.
In particular, for the highly non-ideal mixture (ethanol + water), the Liu model~\cite{Liu2011} cannot capture the pronounced decrease in $D_\mathrm{Ethanol}$ driven by strong polar interactions, while HADES accurately predicts this behavior.

The right column of Figure~\ref{fig_LiuII} shows two mixtures for which experimental values of the self-diffusion coefficients in the pure solvent components are unavailable, so ESE predictions are used instead.
The same picture emerges for these mixtures: although ESE provides adequate predictions for the self-diffusion coefficients in the pure solvents used as input to the Liu model~\cite{Liu2011}, the resulting predictions fail to capture the concentration-dependent trends in the mixtures.
In contrast, HADES accurately predicts the composition dependence for both mixtures (except for one likely outlying data point in the (tetraethylene glycol + water) mixture).
\FloatBarrier

\begin{figure}
\centering
    \includegraphics[width=16cm]{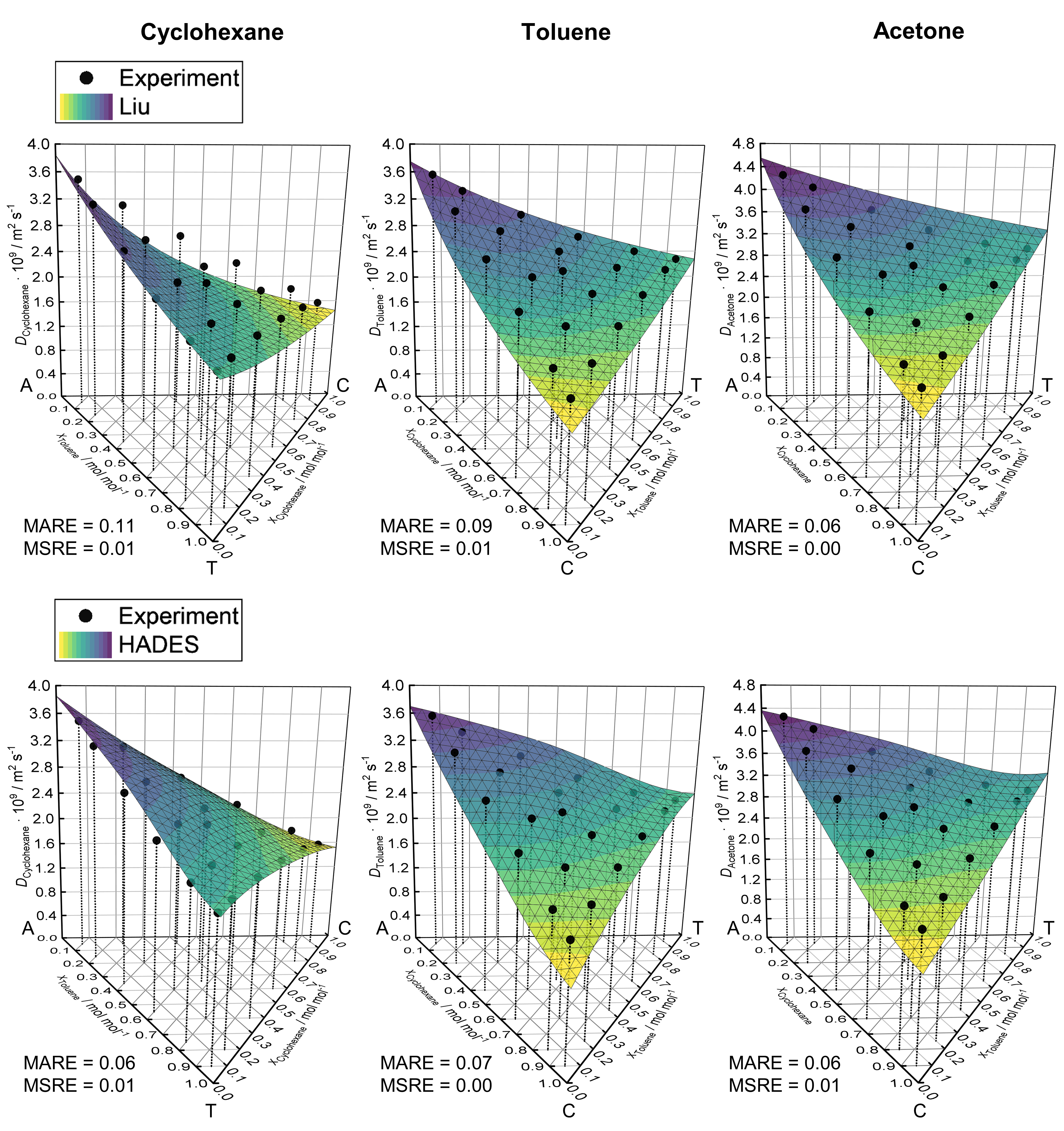}
    \caption{Self-diffusion coefficients at finite concentrations, $D_i$, as a function of composition in a ternary mixture of cyclohexane~(C), toluene~(T), and acetone~(A) at $T=298.15$~K.
    Top row: predictions from the Liu model~\cite{Liu2011}, for which experimental self-diffusion data in the pure solvents were used, except for acetone at infinite dilution in toluene, for which an ESE prediction was used.
    Bottom row: HADES predictions.
    The left, middle, and right columns show the self-diffusion coefficients of cyclohexane, toluene, and acetone, respectively.
    Symbols represent experimental test data, while surfaces denote model predictions.}
    \label{fig_LiuIII}
\end{figure}

Figure~\ref{fig_LiuIII} compares the self-diffusion coefficients at finite concentrations, $D_i$, predicted by the Liu model~\cite{Liu2011} and HADES with the experimental test data for the ternary mixture (cyclohexane + toluene + acetone).
To apply the Liu model~\cite{Liu2011}, we used experimental values throughout, except for the self-diffusion coefficient of acetone at infinite dilution in toluene, for which no experimental data are available, so we used the ESE prediction.

For the self-diffusion coefficient of cyclohexane~(left column), the Liu model~\cite{Liu2011} generally underestimates the experimental data.
HADES yields substantially improved predictions across the full composition range.

For the self-diffusion coefficient of toluene~(middle column), the Liu model~\cite{Liu2011} again underestimates the experimental test data, and HADES provides more accurate predictions.

For the self-diffusion coefficient of acetone~(right column), the Liu model~\cite{Liu2011} and HADES show a comparable predictive performance.
However, HADES better captures the decrease in $D_\mathrm{Acetone}$ at high acetone mole fractions.
\FloatBarrier

The results presented in Figures~\ref{fig_Liu}~--~\ref{fig_LiuIII} demonstrate the improved predictive performance of HADES compared with the Liu model~\cite{Liu2011}.
Together with the results from the previous sections, the results demonstrate that HADES provides compositionally consistent predictions of self-diffusion coefficients for diffusing components at arbitrary concentrations and in mixtures with any number of components.
The chosen deep-set architecture enables HADES to learn from diffusion behavior in pure solvents and extrapolate to mixed solvents, and vice versa, strengthening its predictive performance for both tasks.

\section{Conclusions}

Self-diffusion coefficients are essential for understanding molecular mobility in liquids and for modeling transport processes, yet experimental data remain scarce, particularly for composition-dependent diffusion in mixtures.
Existing prediction approaches are limited in scope, require often-unavailable input data, or fail to describe non-ideal mixture behavior.
HADES closes these gaps.

Building on our hybrid ESE model~\cite{Wagner2026b} for self-diffusion at infinite dilution in pure solvents, HADES extends the framework to arbitrary compositions and multicomponent mixtures through a deep-set architecture that learns a composition-dependent solvent representation.
Trained jointly on pure and mixed solvent data, it provides physically consistent, permutation-invariant predictions across the full composition range, from infinite dilution to pure components, requiring only SMILES strings and pure-component viscosities as input.
For the infinite-dilution state of the diffusing species in mixed solvents, HADES deliberately departs from the Maxwell-Stefan mixing rule of Eq.~\eqref{eq_MS}, trading formal consistency for superior predictive accuracy in non-ideal systems.
The results show this trade-off pays off: HADES matches or outperforms the best available reference models across all classes of self-diffusion coefficients within a single unified framework, without requiring any experimental diffusion data as input.

A particularly notable finding is the bidirectional transfer between pure- and mixed-solvent data: patterns learned from pure-solvent diffusion improve predictions in mixtures, and vice versa.
This suggests that HADES captures the shared physical features of composition-dependent diffusion rather than treating pure and mixed systems as separate regimes, a property that should generalize well beyond the training data.

The current implementation covers organic molecules and water, with molar masses below 1000~g~mol$^{-1}$ and no elements heavier than chlorine, reflecting the scope of available experimental data rather than any inherent limitation of the framework.
Extending HADES to broader chemical spaces remains an important direction for future work, contingent on the availability of high-quality experimental data.

With HADES establishing a reliable and broadly applicable prediction framework for self-diffusion, the natural next step is to extend it to mutual diffusion coefficients at finite concentration, closing the loop between molecular mobility and mixture transport that is central to process design and optimization.

Looking further ahead, the HADES framework offers a promising blueprint for advancing inverse prediction tasks, such as estimating molar masses of unknown components from measured self-diffusion coefficients.
This capability could significantly improve our mixture analysis framework based on nuclear magnetic resonance (NMR) fingerprinting techniques~\cite{Specht2021, Specht2023a, Specht2023b, Specht2023c, Wagner2025, Wagner2026}, which we will explore in future work.

\newpage

\section*{Conflicts of Interest}
There are no conflicts of interest to declare.

\section*{Data Availability}
The final trained HADES model is available on Zenodo at \url{https://doi.org/10.5281/zenodo.21979762}.

\begin{acknowledgement}

We gratefully acknowledge financial support by the Carl Zeiss Foundation in the project 'Halocycles' as well as by DFG in the frame of the Research Training Group 'WERA' (project number 503479768), the Core Facility 'LASE-MR' (project number 537627671) and the Emmy Noether Group of FJ (project number 528649696). Model training was carried out on the high-performance computer Elwetrisch at RPTU under the grant RPTU-MLVT.

\end{acknowledgement}

\begin{suppinfo}

Hyperparameter study; list of diffusing components; list of solvent components.

\end{suppinfo}

\bibliography{achemso-demo}

\end{document}